\documentclass{aipproc}
\layoutstyle{6x9}
\usepackage{epsfig}

\title{Hadrons in Nuclei}
\author{
Ulrich Mosel}{address={Institut fuer Theoretische Physik, Universitaet Giessen\\
D-35392 Giessen, Germany},email={mosel@physik.uni-giessen.de}}

\begin{abstract}
Changes of hadronic properties in dense nuclear matter as
predicted by theory have usually been investigated by means of
relativistic heavy-ion reactions. In this talk I show that
observable consequences of such changes can also be seen in more
elementary reactions on nuclei. Particular emphasis is put on a
discussion of photonuclear reactions; examples are the dilepton
production at $\approx 1$ GeV and the hadron production in nuclei
at 10 - 20 GeV photon energies. The observable effects are
expected to be as large as in relativistic heavy-ion collisions
and can be more directly related to the underlying hadronic
changes.
\end{abstract}

\begin{document}

\maketitle
\setcounter{page}{1}

\section{Introduction}\label{intro}

That hadrons can change their properties and couplings in the
nuclear medium has been well known to nuclear physicists since the
days of the Delta-hole model that dealt with the changes of the
properties of the pion and Delta-resonance inside nuclei
\cite{Ericsson-Weise}. Due to the predominant $p$-wave interaction
of pions with nucleons one observes here a lowering of the pion
branch with increasing pion-momentum and nucleon-density. A direct
observation of this effect is difficult because of the strong
final state interactions (in particular absorption) of the pions.
More recently, experiments at the FSR at GSI have shown that also
the pion rest mass in medium differs from its value in vacuum
\cite{Kienle}. This is interesting since there are also recent
experiments \cite{TAPSsigma} that look for the in-medium changes
of the $\sigma$ meson, the chiral partner of the pion. Any
comparison of scalar and pseudoscalar strength could thus give
information about the degree of chiral symmetry restoration in
nuclear matter.

In addition, experiments for charged kaon production at GSI
\cite{KAOS} have given some evidence for the theoretically
predicted lowering of the $K^-$ mass in medium and the (weaker)
rising of the $K^+$ mass. State-of-the-art calculations of the
in-medium properties of kaons have shown that the usual
quasi-particle approximation for these particles is no longer
justified inside nuclear matter where they acquire a broad
spectral function \cite{Lutz,Tolos}.

At higher energies, at the CERN SPS and most recently at the
Brookhaven RHIC, in-medium changes of vector mesons have found
increased interest, mainly because these mesons couple strongly to
the photon so that electromagnetic signals could yield information
about properties of hadrons deeply embedded into nuclear matter.
Indeed, the CERES experiment \cite{CERES} has found a considerable
excess of dileptons in an invariant mass range from $\approx 300$
MeV to $\approx 700$ MeV as compared to expectations based on the
assumption of freely radiating mesons. This result has found an
explanation in terms of a shift of the $\rho$ meson spectral
function down to lower masses, as expected from theory (see, e.g.,
\cite{Peters,Post,Postneu,Wambach}). However, the actual reason
for the observed dilepton excess is far from clear. Both models
that just shift the pole mass of the vector meson as well as those
that also modify the spectral shape have successfully explained
the data \cite{Cassingdil,RappWam,Rapp}; in addition, even a
calculation that just used the free radiation rates with their --
often quite large -- experimental uncertainties was compatible
with the observations \cite{Koch}. There are also calculations
that attribute the observed effect to radiation from a quark-gluon
plasma \cite{Renk}. While all these quite different model
calculations tend to explain the data, though often with some
model assumptions, their theoretical input is sufficiently
different as to make the inverse conclusion that the data prove
one or another of these scenarios impossible.

I have therefore already some years ago proposed to look for the
theoretically predicted changes of vector meson properties inside
the nuclear medium in reactions on normal nuclei with more
microscopic probes \cite{Hirschegg}. Of course, the nuclear
density felt by the vector mesons in such experiments lies much
below the equilibrium density of nuclear matter, $\rho_0$, so that
naively any density-dependent effects are expected to be much
smaller than in heavy-ion reactions.

On the other hand, there is a big advantage to these experiments:
they proceed with the spectator matter being close to its
equilibrium state. This is essential because all theoretical
predictions of in-medium properties of hadrons are based on an
equilibrium model in which the hadron (vector meson) under
investigation is embedded in cold nuclear matter in equilibrium
and with infinite extension. However, a relativistic heavy-ion
reaction proceeds -- at least initially -- far from equilibrium.
Even if equilibrium is reached in a heavy-ion collision this state
changes by cooling through expansion and particle emission and any
observed signal is built up by integrating over the emissions from
all these different stages of the reaction.

In this talk I summarize results that we have obtained in studies
of observable consequences of in-medium changes of hadronic
spectral functions in reactions of elementary probes with nuclei.
I demonstrate that the expected in-medium sensitivity in such
reactions is as high as that in relativistic heavy-ion collisions
and that in particular photonuclear reactions present an
independent, cleaner testing ground for assumptions made in
analyzing heavy-ion reactions.

\section{Theory}
A large part of the current interest in in-medium properties of
hadrons comes from the hope to learn something about quarks in
nuclei. Indeed, a very simple estimate shows that the chiral
condensate in the nuclear medium is in lowest order in density
given by \cite{Wambach}
\begin{equation}     \label{qbarq}
\langle \bar{q} q \rangle_{\rm med}(\rho,T) \approx \left( 1 -
\sum_h \frac{\Sigma_h \rho^s_h(\rho,T}{f_\pi^2 m_\pi^2} \right)
\langle \bar{q} q \rangle_{\rm vac} ~.
\end{equation}
Here $\rho_s$ is the \emph{scalar} density of the hadron $h$ in
the nuclear system and $\Sigma_h$ the so-called sigma-commutator
that contains information on the chiral properties of $h$. The sum
runs over all hadronic states. While (\ref{qbarq}) is nearly
exact, its actual value is limited because neither the
sigma-commutators of the higher lying hadrons nor their scalar
densities are known. Only at very low temperatures these are
accessible. Here $\rho_s \approx \rho_v \frac{m}{E}$ so that the
condensate drops linearly with the nuclear (vector) density. This
drop can be understood in physical terms: with increasing density
the hadrons with their chirally symmetric phase in their interior
fill in more and more space in the vacuum with its spontaneously
broken chiral symmetry. Note that this is a pure volume effect; it
is there already for a free, non-interacting hadron gas.

How this drop of the scalar condensate translates into observable
hadron masses is not uniquely prescribed. The only rigorous
connection is given by the QCD sum rules that relates an integral
over the hadronic spectral function to a sum over combinations of
quark- and gluon-condensates with powers of $1/Q^2$. It has been
shown \cite{LeupoldMosel} that the QCDSR constrains the hadronic
spectral function, but it does not fix it.

Thus models are needed for the hadronic interactions. The
quantitatively reliable ones can at present be based only on
'classical' hadrons and their interactions. Indeed, in lowest
order in the density the mass and width of an interacting hadron
in nuclear matter at zero temperature and vector density $\rho_v$
are given by (for a meson, for example)
\begin{eqnarray}
{m^*}^2 = m^2 - 4 \pi \Re f_{m N}(q_0,\theta = 0)\, \rho_v
\nonumber \\
m^* \Gamma^* = m \Gamma^0 -  4 \pi \Im f_{mN}(q_0,\theta = 0)\,
\rho_v ~.
\end{eqnarray}
Here $f_{mN}(q_0,\theta = 0)$ is the forward scattering amplitude
for a meson with energy $q_0$ on a nucleon. The width $\Gamma^0$
denotes the free decay width of the particle. For the imaginary
part this is nothing other than the classical relation $\Gamma^* -
\Gamma^0 = v \sigma \rho_v$ for the collision width, where
$\sigma$ is the total cross section. This can easily be seen by
using the optical theorem.

Note that such a picture also encompasses the change of the chiral
condensate in (\ref{qbarq}), obtained there for non-interacting
hadrons. If the spectral function of a non-interacting hadron
changes as a function of density, then in a classical hadronic
theory, which works with fixed free hadron masses, this change
will show up as an energy-dependent interaction and is thus
contained in any empirical phenomenological cross section.

\section{Dilepton Production}

\begin{figure}[h]
\centering{\includegraphics[width=9cm]{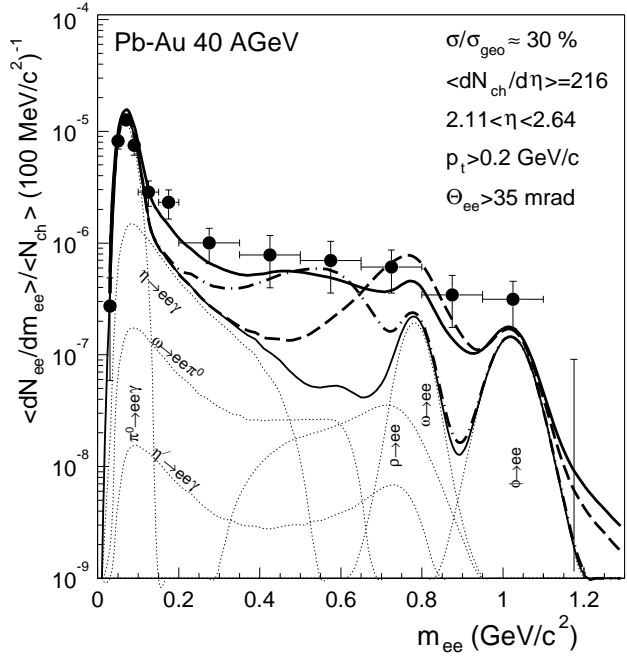}}
 \vspace*{-0.2cm} \caption{Invariant  dilepton mass spectrum
 obtained with the CERES experiment in Pb + Au collisions at 40
 AGeV (from \cite{CERES}). The thin curves give the contributions
 of individual hadronic sources to the total dilepton yield, the
 fat solid (modified spectral function) and the dash-dotted
 (dropping mass only) curves give the results of calculations
 \cite{Rapp} employing an in-medium modified spectral function of the vector
 mesons.} \label{CERES}
\end{figure}

Dileptons, i.e. electron-positron pairs, in the outgoing channel
are an ideal probe for in-medium properties of hadrons since they
-- in contrast to hadronic probes -- experience no strong final
state interaction. A first experiment to look for these dileptons
in heavy-ion reactions was the DLS experiment at the BEVALAC in
Berkeley \cite{DLS}. Later on, and in a higher energy regime, the
CERES experiment has received a lot of attention for its
observation of an excess of dileptons with invariant masses below
those of the lightest vector mesons \cite{CERES}. Explanations of
this excess have focussed on a change of in-medium properties of
these vector mesons in dense nuclear matter (see e.g.\
\cite{Cassingdil,RappWam}). The radiating sources can be nicely
seen in Fig.~\ref{CERES} that shows the dilepton spectrum obtained
in a low-energy run at 40 AGeV together with the elementary
sources of dilepton radiation.

The figure exhibits clearly the rather strong contributions of the
vector mesons -- both direct and through their Dalitz decay -- at
invariant masses above about 500 MeV. If this strength is shifted
downward, caused by an in-medium change of the vector-meson
spectral functions, then the observed excess can be explained as
has been shown by various authors (see e.g.\ \cite{Brat-Cass} for
a review of such experiments).

As mentioned above such explanations always suffer from an
inherent inconsistency: while the observed signal integrates over
many different stages of the collision -- nonequilibrium and
equilibrium, the latter at various densities and temperatures --
the theoretical input is always calculated under the assumption of
a vector meson in nuclear matter in equilibrium. We have therefore
looked for possible effects in reactions that proceed much closer
to equilibrium and have thus studied the dilepton production in
reactions on nuclear targets involving more elementary
projectiles. It is not \emph{a priori} hopeless to look for
in-medium effects in ordinary nuclei: Even in relativistic
heavy-ion reactions that reach baryonic densities of the order of
3 - 10 $\rho_0$ many observed dileptons actually stem from
densities that are much lower than these high peak densities.
Transport simulations have shown \cite{Brat-Cass} that even at the
CERES energies about 1/2 of all dileptons come from densities
lower than $2 \rho_0$. This is so because in such reactions the
pion-density gets quite large in particular in the late stages of
the collision, where the baryonic matter expands and its density
becomes low again. Correpondingly many vector mesons are formed
(through $\pi + \pi -> \rho$) late in the collision and their
decay to dileptons thus happens at low baryon densities.

It is thus a quantitative question if any observable effects of
in-medium changes of hadronic properties survive if the densities
probed are always $\le \rho_0$. With the aim of answering this
question we have over the last few years undertaken a number of
calculations for proton- \cite{Bratprot}, pion-
\cite{Weidmann,Effepi} and photon- \cite{Effephot} induced
reactions. All of them have one feature in common: they treat the
final state incoherently in a coupled channel transport
calculation that allows for elastic and inelastic scattering of,
particle production by  and absorption of the produced vector
mesons. We have also looked into the prospects of using reactions
with hadronic \cite{MuehlPhi} final states. In this case, the
photoproduction of $\phi$ mesons on nuclei, our conclusion was
that no in-medium signal from the $\phi$ could be observed due to
the strong final state interactions of the outgoing kaons, the
decay products of the $\phi$ meson. A semi-hadronic final state,
such as $\pi^0\gamma$, as obtained in the photoproduction of
$\omega$ mesons on nuclei looks more promising \cite{MuehlOm}.

All the photonuclear calculations are done in a combination of
coherent initial state interactions that lead to shadowing at
photon energies above about 1 GeV and incoherent final state
interactions. The shadowed incoming photon produces, for example,
a vector meson which then cascades through the nucleus. The latter
process we describe by means of a coupled-channel transport
theory. The details are discussed in ref. \cite{Effe}. A new
feature of these calculations is that vector mesons with their
correct spectral functions can actually be produced and
transported consistently. This is quite an advantage over earlier
treatments \cite{Brat-Cass} in which the mesons were always
produced and transported with their pole mass and their spectral
function was later on folded in only for their decay.

\begin{figure}[h]
\vspace*{-0.5cm}
\centering{\epsfig{figure=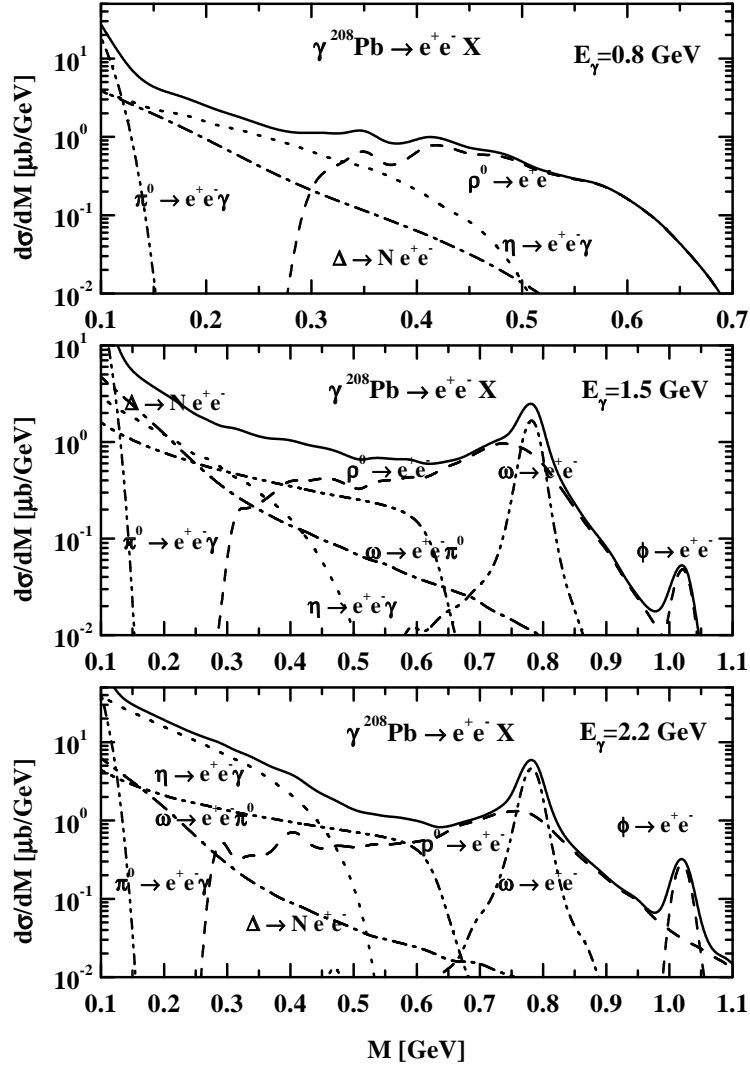,width=12cm}} \vspace*{-2cm}

\caption{Hadronic contributions to dilepton invariant mass spectra
for $\gamma + ^{208}Pb$ at the three photon energies given (from
\cite{Effephot}). Compare with Fig.~\ref{CERES}.} \label{Fige+e-}
\end{figure}

\begin{figure}[htb]
\centering{\includegraphics[width=13cm]{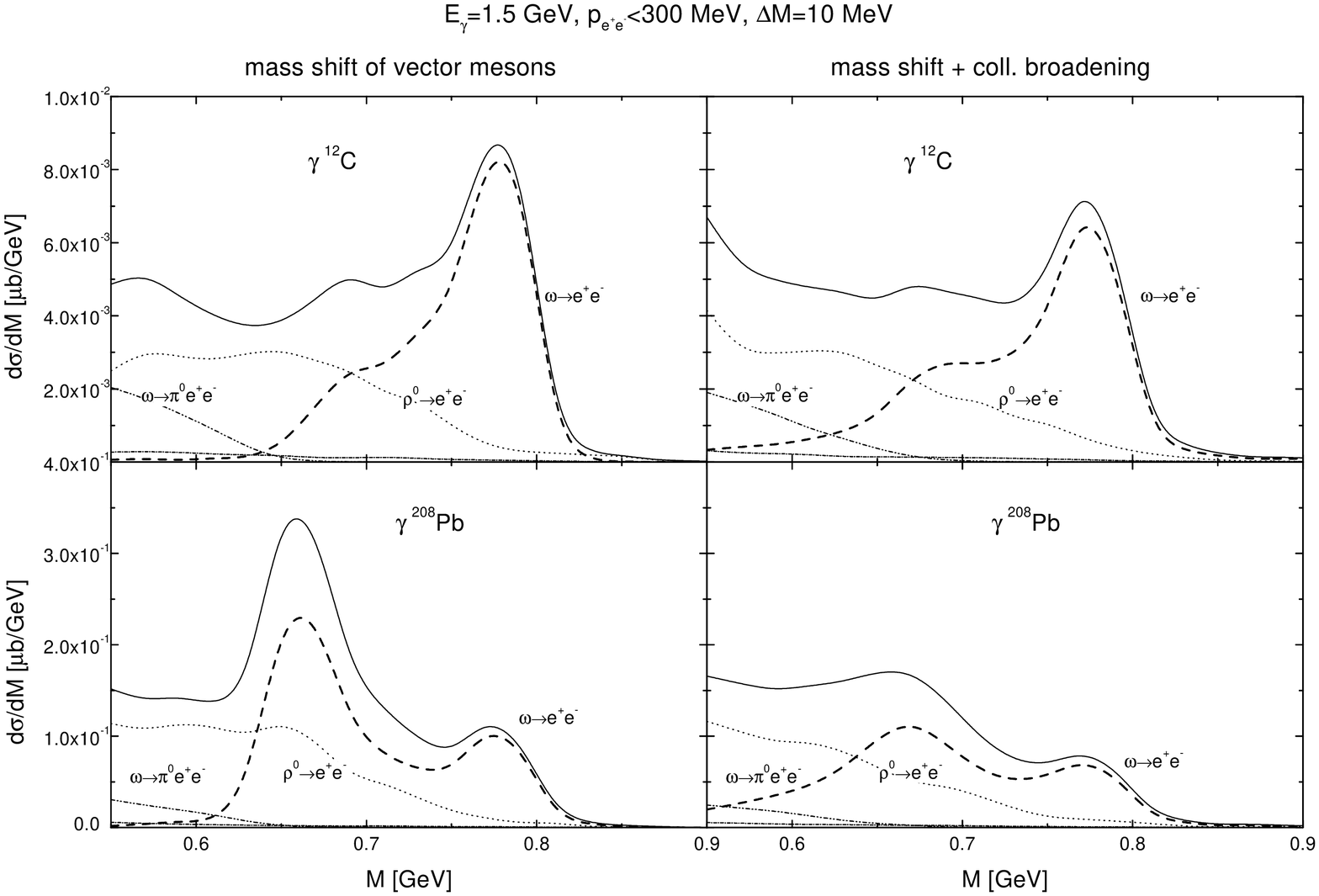}}
\vspace*{-0.5cm} \caption{Dilepton mass yield with a
dilepton-momentum cut of 300 MeV. Shown on the left are results of
a calculation that uses only a shift of the pole mass of the
vector mesons. On the right, results are given for a calculation
using both mass shift and collisional broadening (from
\cite{Effe}).} \label{Figdlim}
\end{figure}

A typical result of such a calculation for the dilepton yield --
after removing the Bethe-Heitler component -- is given in Fig.
\ref{Fige+e-}.
Comparing this figure with Fig. \ref{CERES} shows that in a
photon-induced reaction at 1 - 2 GeV photon energy exactly the
same sources, and none less, contribute to the dilepton yield as
in relativistic heavy-ion collisions at 40 AGeV! The question now
remains if we can expect any observable effect of possible
in-medium changes of the vector meson spectral functions in medium
in such an experiment on the nucleus where -- due to surface
effects -- the average nucleon density is below $\rho_0$. This
question is answered, for example, by the results of Fig.\
\ref{Figdlim}.
This figure shows the dilepton spectra to be expected if a
suitable cut on the dilepton momenta is imposed; with this cut
slow vector mesons are enriched. In the realistic case shown on
the right, which contains both a collision broadening and a mass
shift, it is obvious that a major signal is to be expected: in the
heavy nucleus $Pb$ the $\omega$-peak has completely disappeared
from the spectrum. The sensitivity of such reactions is thus as
large as that observed in ultrarelativistic heavy-ion reactions.

An experimental verification of this prediction would be a major
step forward in our understanding of in-medium changes\footnote{An
experiment at JLAB is under way \cite{Weygand}.}. It would
obviously present a purely hadronic base-line to all data on top
of which all 'fancier' explanations of the CERES effect in terms
of radiation from a QGP and the such would have to live.

\section{Hadron Formation}

An in-medium effect different from the ones discussed so far
happens when particles are produced by high-energy projectiles
inside a nuclear medium. Then measurements of their yield can
actually give information on the time it takes until the newly
created particle has evolved into a 'normal', fully interacting
particle. The longer this formation time is, the less absorption
will take place. Thus nuclei serve as a kind of 'microdetector'
for the determination of formation times. Obviously, this
phenomenon is closely related to color transparancy. A
particularly appealing probe are jets that emerge from the nuclei.

A major experimental effort at RHIC experiments has gone into the
observation of jets in ultrarelativistic heavy-ion collisions and
the determination of their interaction with the surrounding quark
or hadronic matter \cite{Jet}. Such experiments are obviously very
sensitive to hadron formation times. In addition they can yield
information on interactions while the final hadron is still being
formed.

A complementary process is given by the latest HERMES results at
HERA for photon-induced hadron production at high energies
\cite{Hermes}. Here the photon-energies are of the order of 10 -
20 GeV, with rather moderate $Q^2 \approx 2$ GeV$^2$. Again, the
advantage of such experiments is that the nuclear matter with
which the interactions happen is at rest and in equilibrium.

In the high-energy regime the shadowing of the incoming photon,
which is due to interference between interactions of the incoming
bare photon and its hadronic components with the nucleons, becomes
important. This coherence in the incoming state has to be combined
with the incoherent treatment of the final state interactions in
transport theory. For this purpose T.\ Falter has derived a novel
expression for incoherent particle production on the nucleus
\cite{falterinc} that allows for a clean-cut separation of the
coherent initial state and the incoherent final state interactions
which we again treat with our coupled-channel transport theory.
\begin{figure}[h]
\vspace*{-0.0cm}
 \centering{\includegraphics[width=10.0cm]{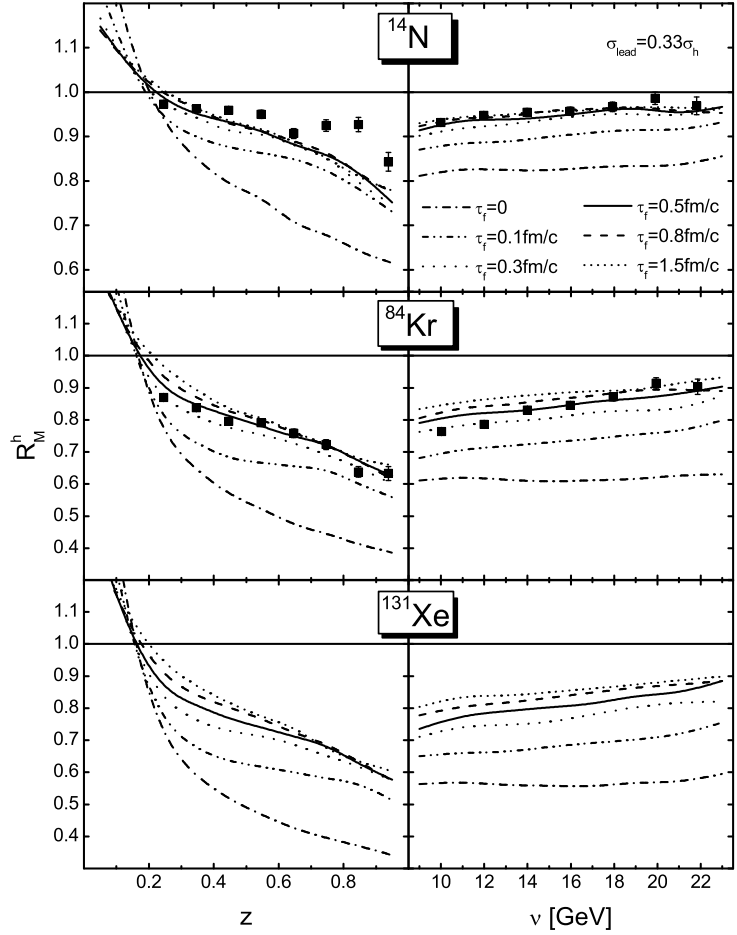}}
\vspace*{-0.5cm}

\caption{Multiplicity of produced hadrons normalized to the
deuterium as a function of photon-energy $\nu$ (right) and of the
energy of the produced hadrons relative to the photon-energy, $ z
= E_h/\nu$. The curves are calculated for different formation
times given in the figure (from \cite{falterform}).}
\label{formtime}
\end{figure}

An example of the results obtained is given in Fig.\
\ref{formtime}. The figure clearly shows that the observed hadron
multiplicities can be described only with formation times $>
\approx 0.3$ fm. The curves obtained with larger formation times
all lie very close together. This is a consequence of the finite
size of the target nucleus: if the formation time is larger than
the time needed for the preformed hadron to transverse the
nucleus, then the sensitivity to the formation time is lost. In
\cite{falterform} we have also shown that the $z$-dependence on
the left side exhibits some sensitivity to the interactions of the
leading hadrons during the formation; the curves show in Fig.\
\ref{formtime} are obtained with a leading hadron cross section of
$0.33 \sigma_h$, where $\sigma_h$ is the 'normal' hadronic
interaction cross section.

\section{Conclusions}\label{concl}
In this talk I have illustrated with the help of two examples that
photonuclear reactions can yield information that is important and
relevant for an understanding of high density--high temperature
phenomena in ultrarelativistic heavy-ion collisions. I have shown
that the expected sensitivity of dilepton spectra in photonuclear
reactions in the 1 - 2 GeV range is as large as that in
ultrarelativistic heavy-ion collisions. I have also illustrated
that the analysis of hadron production spectra in high-energy
electroproduction experiments at HERMES gives information about
the interaction of forming hadrons with the surrounding hadronic
matter. This is important for any analysis that tries to obtain
signals for a QGP by analysing high-energy jet formation in
ultrarelativistic heavy-ion reactions.

\section*{Acknowledgement}
Large parts of this talk are based on work of two of my students,
Martin Effenberger and Thomas Falter. This work has been supported
by the Deutsche Forschungsgemeinschaft, the BMBF and GSI
Darmstadt.

\vfill\eject
\end{document}